\documentclass[fleqn,usenatbib]{mnras}

\usepackage{mathrsfs}
\usepackage{graphicx}
\usepackage{subfigure}
\usepackage{epstopdf}
\usepackage{amsmath}
\usepackage{amssymb}
\usepackage{hyperref}
\usepackage{color}

\DeclareRobustCommand{\VAN}[3]{#2}
\let\VANthebibliography\thebibliography
\def\thebibliography{\DeclareRobustCommand{\VAN}[3]{##3}\VANthebibliography}

\title[Constraints on compact DM from lensing of GWs]{Constraints on compact dark matter from lensing of gravitational waves for the third-generation gravitational wave detector}

\author[H. Zhou et al. 2022]{
Huan Zhou,$^{1}$
Zhengxiang Li,$^{2}$\thanks{E-mail: zxli918@bnu.edu.cn}
Kai Liao,$^{3}$
and Zhiqi Huang$^{1}$
\\
$^{1}$School of Physics and Astronomy, Sun Yat-sen University, Zhuhai, 519082, China\\
$^{2}$Department of Astronomy, Beijing Normal University, Beijing 100875, China\\
$^{3}$School of Physics and Technology, Wuhan University, Wuhan 430072, China}
\date{Accepted 2022 October 10. Received 2022 October 8; in original form 2022 August 9}

\pubyear{2022}

\begin{document}
\label{firstpage}
\pagerange{\pageref{firstpage}--\pageref{lastpage}}
\maketitle

\begin{abstract}
Since the first gravitational wave (GW) event from binary black hole (BBH) was detected by LIGO-Virgo, GWs have become a useful probe on astrophysics and cosmology. If compact dark matter (DM) objects e.g. primordial black holes, contribute a significant fraction of dark matter at wide mass range, they will cause microlensing in the GW signals with long wavelengths that are distinct from the lensing effects of electromagnetic signals from astrophysical objects. In this paper, we apply the lensing effect of GW from BBH to derive constraints on the abundance of compact DM for the Cosmic Explorer, a third-generation ground-based GW detector. We firstly consider two channels of formation of BBH that contribute to low and high redshift GW sources, including the astrophysical origin BBH scenario, and the primordial origin BBH scenario. Secondly, comparing with the method of optical depth, we use the Bayesian analysis to derive constraints on the abundance of compact DM with different mass function of lens taken into consideration. For a null search with $1000$ detected GW events of BBH, we find that the abundance of compact DM could be constrained to $\lesssim0.1\%$ in the mass range $\geq500~M_{\odot}$ at $68\%$ confidence level. In addition, if a GW event lensed by a compact DM object with $M_{\rm l}\in[100~M_{\odot},300~M_{\odot}]$ is detected in $100$ detected GW events of BBH, we can derive that the estimation of the abundance of compact DM is from $2.3\%$ to $25.2\%$ in this mass range with the Bayesian analysis.
\end{abstract}

\begin{keywords}
(cosmology): dark matter, gravitational lensing: micro,  gravitational waves
\end{keywords}

\section{Introduction}\label{sec1}
The first detection of gravitational wave (GW) event from binary black hole (BBH) merger opened a new window in astronomy~\citep{Abbott2016}. GWs have been proposed as very powerful probes of astrophysics and cosmology, such as solving Hubble tension~\citep{Hotokezaka2019, Feeney2021}, constraining  the speed of GWs~\citep{Abbott2017}, and constraints on the dark matter candidates~\citep{Yoshida2018, Basak2022}. With the increasing sensitivity of the current and future detectors, such as Einstein Telescope (ET)~\citep{Reitze2019}, Cosmic Explorer (CE)~\citep{Reitze2019}, Laser Interferometer Space Antenna (LISA)~\citep{Amaro-Seoane2017}, Taiji~\citep{Hu2017}, Tianqin~\citep{Luo2016}, DECihertz Interferometer Gravitational wave Observatory (DECIGO)~\citep{Kawamura2019}, Atom Interferometer Observatory and Network (AION)~\citep{Badurina2020}, Mid-band Atomic Gravitational Wave Interferometric Sensor (MAGIS)~\citep{Graham2017}, and Atomic Experiment for Dark Matter and Gravity Exploration (AEDGE)~\citep{El-Neaj2020}, the increasingly detectable GW events and multiband GW observations are expected to solve these open questions in modern astrophysics and cosmology.

Dark matter (DM) makes up about one quarter of the total energy budget of the universe, which is consistent with many popular cosmological observations. However, we still know little about the constituent of DM, especially in small scales. Theoretical models from particle physics and cosmology predict many types of compact DM which include the massive compact halo objects (MACHOs), primordial black holes (PBHs), axion mini-clusters, compact mini halos, boson stars, fermion stars and so on. In particular, PBH is taken as a potential candidate of DM which has been a field of great astrophysical interest.(See~\citet{Sasaki2018,Green2020,Carr2021} for recent reviews.) In theory, the mass of compact DM can range from the Planck mass to the level of the black hole in the centre of the galaxy. Therefore, numerous methods have been proposed to constrain the fraction of compact DM in DM $f_{\rm DM}$ at present universe in various possible mass windows. Gravitational lensing effect is a powerful probe to constrain the abundance of compact DM over a broad mass range from $\mathcal{O}(10^{-10}~M_{\odot})$ to $\mathcal{O}(10^{10}~M_{\odot})$. For example, observing a large number of stars and looking for amplifications in their brightness caused by lensing effect of intervening massive objects could yield constraints on the abundance of deflectors~\citep{Allsman2001,Tisserand2007,Griest2013,Niikura2019a, Niikura2019b}. Searching lensing systems that observed signals of transient sources like the fast radio bursts (FRBs) or gamma ray bursts (GRBs) would appear echoes were proposed to put constraints on the compact DM~\citep{Munoz2016, Ji2018,Laha2020,Liao2020b,Zhou2022a,Zhou2022b,Lin2022}. Searching multiple images produced by milli-lensing of possible persistent sources like the compact radio sources (CRSs) is another powerful probe~\citep{Kassiola1991,Wilkinson2001,Zhou2022c}. Similar to electromagnetic waves, GWs can also be lensed by compact DM. However, the most significant difference when looking for the lensing of GWs from compact object binaries instead of lensing of stellar light is that the characteristic wavelength of GW signals is very long and wave optics effect become important. Therefore, we should consider the diffraction effect in the lensing of GW, which distorts the GW waveform as the fringes. \citet{Jung2019} first proposed that the lensing effect of GWs observed by advanced LIGO can probe the compact DM. There are many subsequent studies using the lensing effect of GW have been performed~\citep{Liao2020a,Urrutia2021,Wang2021,Basak2022}. Especially in \citet{Basak2022} work, they constrain the fraction of compact DM in the mass range $10^2-10^5~M_{\odot}$ to be less than $50\%-80\%$ from the non-detection of microlensing signals in currently available BBH events detected by LIGO-Virgo. This result is still weaker compared with the results of other methods~\citep{Sasaki2018,Green2020,Carr2021}. 

In this paper, based on GW events that mainly consist of signals from the coalescence of BBH, we study the ability of  the Cosmic Explorer (CE), a future third-generation ground-based GW detector, to constrain the abundance of compact DM through gravitational lensing of GW signals from BBH~\footnote{For the third generation ground-based GW detectors, the sensitivities of ET and CE are approximately the same, which determines their abilities to constrain the abundance of compact DM are similar. Therefore, we only use CE as the representative of the third generation ground-based GW detectors.}. It should be noted that, there are two most important differences between our analyses and previous works~\citep{Jung2019,Liao2020a, Urrutia2021,Wang2021}. Firstly, there are still many discussions about the origin mechanism of GWs of BBH. Therefore, we assume two channels of formation of BBH, including astrophysical and primordial origins, which correspond to different redshift and mass distribution of GW sources. Secondly, based on~\citet{Basak2022}, we compare the constraints of the abundance of compact DM from the Bayesian analysis with the results from the method of optical depth. Moreover, previous works are based on the monochromatic mass distribution of lens, here we add the extended mass distribution of lens into the method of Bayesian analysis to constrain the compact DM.

This paper is organized as follows: In Section~\ref{sec2}, we introduce the lensing effect of GW with wave optical description; In Section~\ref{sec3}, we firstly give our simulated models, the primordial origin BBH model and the astrophysical origin BBH model. Then we introduce the method of Bayesian analysis. The results of constraints are presented in Section~\ref{sec4}. Finally, we summarize our conclusions in Section~\ref{sec5}. Throughout, we use the concordance $\Lambda$CDM cosmology with the best-fitting parameters from the recent $Planck$ observations~\citep{Planck2018}, and the natural units of $G=c=1$ in all equations.

\section{Lensing of gravitational wave}\label{sec2}
In this Section, we briefly review the theory of lensing of GWs described by the wave optics limit~\citep{Takahashi2003}. A GW is described by a tiny perturbation $h_{\mu\nu}$ over the background metric $g_{\mu\nu}$. The amplitude $h$ of the perturbation $h_{\mu\nu}$ follows the equation as
\begin{equation}\label{eq1}
\partial_{\mu}(\sqrt{-g}g^{\mu\nu}\partial_{\nu}h)=0.
\end{equation}
Based on the lesning events occurring near the deflector, the equation of lensing GW can be approximated in the Fourier space as
\begin{equation}\label{eq2}
(\Delta+4\pi^2f^2)\bar{h}=16\pi^2f^2U\bar{h},
\end{equation}
where $U$ is the lensing potential in the background metric $g_{\mu\nu}$, and $\bar{h}$ is Fourier transform of $h$. We can define the dimensionless amplification factor as
\begin{equation}\label{eq3}
F(f)=\frac{\bar{h}^{\rm L}(f)}{\bar{h}_0(f)},
\end{equation}
where $\bar{h}^{\rm L}(f)$ and $\bar{h}_0(f)$ correspond to lensed and unlensed ($U=0$) signal, respectively. For the unlensed waveform $\bar{h}_0(f)$ from BBHs merger event, we ignore higher order post-Newtonian terms for simplicity~\citep{Thorne1987},
\begin{equation}\label{eq4}
\bar{h}_0(f)=\sqrt{\frac{5}{24}}\frac{\mathcal{M}_z^{5/6}\mathcal{F}}{\pi^{2/3}d_{\rm L}(z)}f^{-7/6}\exp\{i[\Psi(f)+\Psi_0]\},
\end{equation}
where
\begin{equation}\label{eq5}
\Psi(f)=2\pi ft_{\rm c}+\frac{3}{128}(\pi\mathcal{M}_zf)^{-5/3},
\end{equation}
$\mathcal{M}_z$, $t_{\rm c}$ and $d_{\rm L}$ represent the redshifted chirp mass, coalescence time and luminosity distance ,respectively. $\mathcal{F}$ is the angular orientation function which contains all angle dependence of the detector response to GW events, and we give a distribution for $\mathcal{F}$~\citep{Finn1996}
\begin{equation}\label{eq6}
P(\mathcal{F})=20\mathcal{F}(1-\mathcal{F})^3\mathcal{H}(\mathcal{F})\mathcal{H}(1-\mathcal{F}),
\end{equation}
where $\mathcal{H}$ is the Heaviside step function. The solution of Equation~(\ref{eq3}) in the thin lens approximation is given by
\begin{equation}\label{eq7}
F(f)=\frac{D_{\rm s}D_{\rm l}\theta_{\rm E}^2(1+z_{\rm l})f}{D_{\rm ls}i}\int d^2x~\exp[2\pi i ft_{\rm d}(\vec{x},\vec{y})],
\end{equation}
where $D_{\rm s}$, $D_{\rm l}$ and $D_{\rm ls}$ represent the angular diameter distance to the source, to the lens, and between the source and the lens, respectively. In addition, $\theta_{\rm E}$ is Einstein radius, and $\vec{y}$ is the impact position of source. $t_{\rm d}$ is the arrival time delay of the wave to the observer
\begin{equation}\label{eq8}
t_d(\vec{x},\vec{y})=\frac{D_{\rm s}D_{\rm l}\theta_{\rm E}^2(1+z_{\rm l})}{D_{\rm ls}}\bigg[\frac{1}{2}|\vec{x}-\vec{y}|^2-\psi(\vec{x})-\phi_{\rm m}(\vec{y})\bigg],
\end{equation}
where $\psi(\vec{x})$ is dimensionless deflection potential and $\phi_{\rm m}(\vec{y})$ is chosen such that the minimum arrival time delay is zero. For a point mass lens, we can use the rescaled position $x=|\vec{x}|$, $y=|\vec{y}|$ to get potential $\psi(\vec{x})=\ln x$ and $\phi_m(\vec{y})=0.5(x_m-y)^2-\ln x_m$ with $x_m=0.5(y+\sqrt{y^2+4})$. In this case, $F(f)$ can be written as
\begin{equation}\label{eq9}
\begin{split}
F(f)=\exp\bigg[\frac{\pi\omega}{4}+i\frac{\omega}{2}\bigg(\ln \frac{\omega}{2}-2\phi_{\rm m}(y)\bigg)\bigg]\\
\times\Gamma\bigg(1-\frac{i\omega}{2}\bigg){_{1}F_1}\bigg(i\frac{\omega}{2},1,\frac{i}{2}\omega y^2\bigg),
\end{split}
\end{equation}
where $\omega=8\pi M_{\rm l}(1+z_{\rm l})f$ is a dimensionless parameter and $_{1}F_1$ is the confluent hypergeometric function. For a point mass lens, the dimensionless amplification factor $F(f)$ depends only on the redshifted mass of lens $M_{\rm l}^{z}\equiv(1+z_{\rm l})M_{\rm l}$ and the source position $y$.

\section{Methodology}\label{sec3}
In this Section, we introduce the simulation of GWs and specify the Bayesian analysis.

\subsection{Simulation}\label{sec31}
\begin{figure}
    \centering
     \includegraphics[width=0.45\textwidth, height=0.4\textwidth]{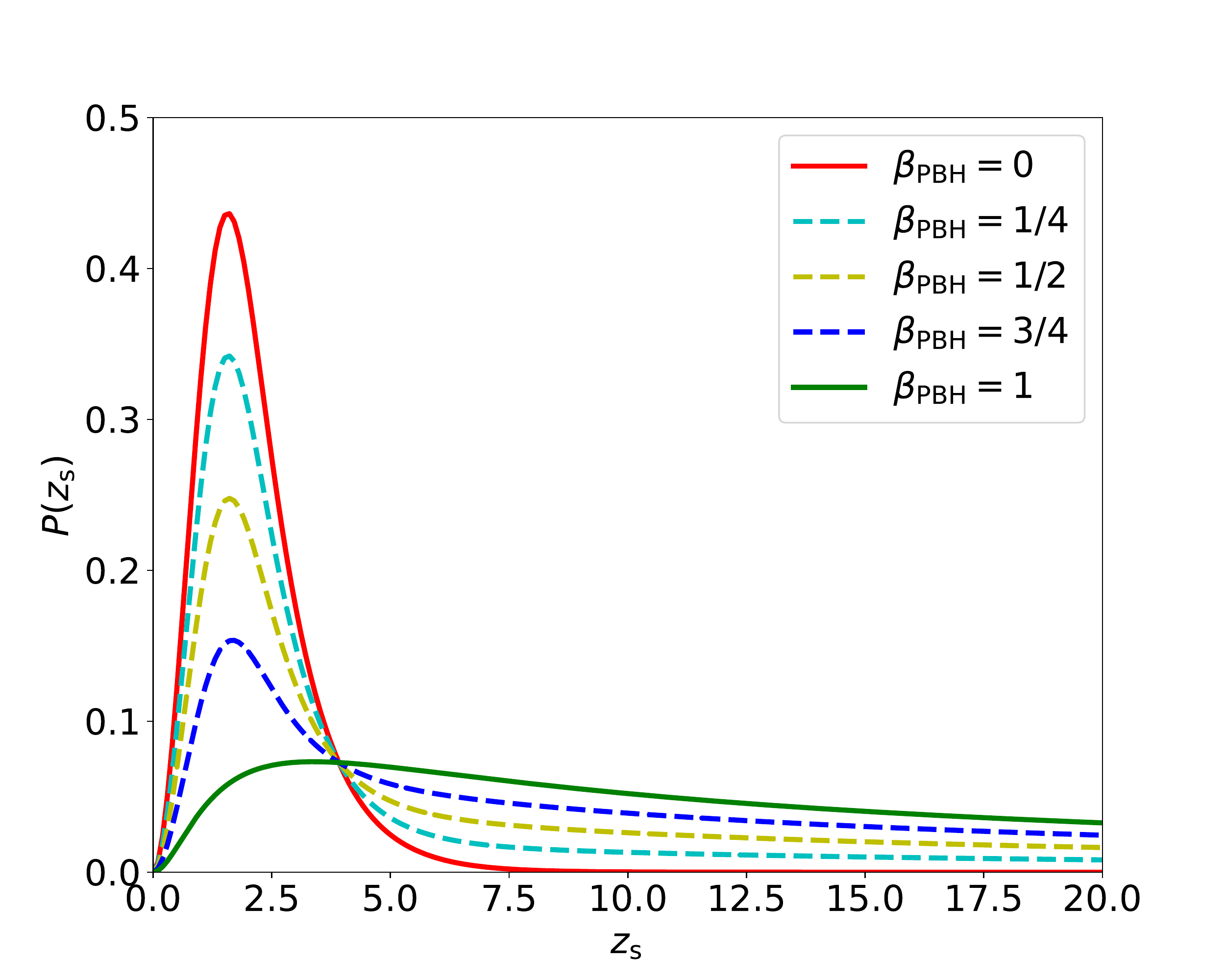}
     \caption{Redshift distribution of GW events of BBH with different $\beta_{\rm PBH}$.}\label{fig1}
\end{figure}

Understanding the origin of GWs of BBHs is an essential scientific goal and is still an open issue. There are two explanations for observed BBH events, i.e. the PBH binary model and the astrophysical black hole (ABH) binary model.  

For PBH binaries, PBHs formed in the early universe from gravitational collapse of primordial density perturbations~\citep{Sasaki2018,Green2020,Carr2021}, and have quite different evolutionary histories than ABHs which originate from the demise of massive stars. There are two distinct formation mechanisms of PBH binaries which form at different epoch in the cosmic history. The first mechanism of PBH binaries operates by decoupling from the cosmic expansion in the universe dominated by radiation~\citep{Sasaki2016, Haimoud2017, Chen2018, Raidal2017, Raidal2018}. The second mechanism is that PBHs binaries form in the late universe by the close encounter~\citep{Sasaki2016,Raidal2017}. Comparing with the second formation mechanism of PBH binaries,  the mergers of PBH binaries from the first formation mechanism contributes dominant GW sources of BBHs~\citep{Sasaki2016,Raidal2017}. Therefore, as suggested in ~\citet{Haimoud2017, Chen2018, Raidal2017, Raidal2018}, we mainly consider the first formation mechanism of PBH binaries, and the comoving merger rate density in units of $\rm Gpc^{-3}yr^{-1}$ is given as
\begin{equation}\label{eq10}
R_{\rm PBH}(z)=R_{\rm PBH,0}\bigg(\frac{t(z)}{t_0}\bigg)^{-\frac{34}{37}},
\end{equation}
where $R_{\rm PBH,0}$ and $t_0$ represent the $R_{\rm PBH}(z=0)$ and the current age of the universe, respectively. Based on the merger rate, we can obtain the normalized redshift distribution of PBH binaries as
\begin{equation}\label{eq11}
P_{\rm PBH}(z)=\frac{1}{Z_{\rm PBH}}\frac{R_{\rm PBH}(z)}{1+z}\frac{dV_{\rm c}}{dz},
\end{equation}
where $Z_{\rm PBH}$ is the normalization constant and $dV_{\rm c}/dz$ is the differential comoving volume. The term $1+z$ accounts for the cosmological time dilation. Since we still do not understand the detailed process of PBH formation, we first assume that PBH mass distribution in PBH binaries is described by the following model-independent log-normal mass function
\begin{equation}\label{eq12}
P_{\rm PBH}(m)=\frac{1}{\sqrt{2\pi}\sigma m}\exp\bigg(-\frac{\ln^2(m/m_{\rm c})}{2\sigma^2}\bigg),
\end{equation}
where $m_{\rm c}$ and $\sigma$ denote the peak mass of $mP(m)$ and the width of mass spectrum, respectively. This model-independent parameterization is often a good approximation if PBHs produced from a smooth symmetric peak in the inflationary power spectrum~\citep{Green2016,Kannike2017}, and is often used in the literature to derive constraints on the PBH from GW measurements detected by the LIGO-Virgo~\citep{Chen2019,Wu2020,Luca2020,Gert2021,Kaze2021}.

For ABH binaries, we follow the widely accepted ``Vangioni" model to estimate the comoving merger rate~\citep{Dvorkin2016}, which is a convolution of the birthrate of ABHs $R_{\rm b}(t(z)-t, m)$ with the distribution of the time delays $P(t)$ between the formation and merger
\begin{equation}\label{eq13}
R_{\rm ABH}(z)=\int_{>1M_{\odot}}\int_{t_{\rm min}}^{t_{\rm max}}R_{\rm b}(t(z)-t, m)\times P(t)dtdm,
\end{equation}
where $t_{\rm min}=50~\rm Myr$, and $t_{\rm max}$ is set to the Hubble time. In addition, $R_{\rm b}(t, m)$ can be estimated by~\citep{Dvorkin2016}
\begin{equation}\label{eq14}
R_{\rm b}(t, m_{\rm bh})=\int\psi[t-\tau(m')]\phi(m')\delta(m'-g^{-1}_{\rm b}(m_{\rm bh}))dm',
\end{equation}
where $m_{\rm bh}$ is the mass of the remnant black hole, $\tau(m)$ is the lifetime of a progenitor star and $\phi(m)$ is the initial mass function. The star formation rate $\psi(t)$ in $R_{\rm b}(t, m_{\rm bh})$ is given by
\begin{equation}\label{eq15}
\psi(z)=k\frac{a~\exp[b(z-z_m)]}{a-b+b~\exp[a(z-z_m)]},
\end{equation}
where the parameter set $\{k,a,b,z_m\}$ is from the Fiducial model $\{0.178~M_{\odot}\rm{yr}^{-1}\rm{Mpc}^{-3}, 2.37, 1.80, 2.00\}$ given in~\citet{Vangioni2015}, which corresponds to the classical isolated binary evolution of ABH binaries model. Similar to the PBH binaries, the redshift distribution of ABH binaries can be given as
\begin{equation}\label{eq16}
P_{\rm ABH}(z)=\frac{1}{Z_{\rm ABH}}\frac{R_{\rm ABH}(z)}{1+z}\frac{dV_{\rm c}}{dz},
\end{equation}
where $Z_{\rm ABH}$ is the normalization constant. We use a power law mass distribution for the heavier black hole in the ABH binaries as~\citep{Liao2020a,Basak2022}
\begin{equation}\label{eq17}
P_{\rm ABH}(m)=\frac{1}{Z_{\rm m}}m^{-2.35}\mathcal{H}(m-5~M_{\odot})\mathcal{H}(100~M_{\odot}-m),
\end{equation}
where $Z_{\rm m}$ is the normalization constant.

Next, we consider the redshift distribution of simulated GW events of BBH include both PBH and ABH populations
\begin{equation}\label{eq18}
P(z)=\beta_{\rm PBH}P_{\rm PBH}(z)+(1-\beta_{\rm PBH})P_{\rm ABH}(z),
\end{equation}
where $\beta_{\rm PBH}$ is the fraction of the PBH merger rate to the total merger rate. As shown in Figure~\ref{fig1}, when $\beta_{\rm PBH}=0$ and $\beta_{\rm PBH}=1$, the redshift distribution of GW events of BBH recovers to $P_{\rm ABH}(z)$ and $P_{\rm PBH}(z)$, respectively. In particular, the contribution to the number of BBH events from high redshift ($z>7$) for ABH binaries model can be negligible compared with the PBH binaries model. If the range of fraction of PBH in DM at present universe can be from $10^{-3}$ to $10^{-2}$, all the merging models of PBH binaries can provide enough BBH events to explain the case of $\beta_{\rm PBH}=1$ theoretically~\citep{Sasaki2016, Haimoud2017, Chen2018, Raidal2017, Raidal2018}. Furthermore, possibilities of the above merging models have been explored by combining the PBH population with log-normal mass function and GW events detected by LIGO-Virgo  \citep{Chen2019,Wu2020,Luca2020,Kaze2021}. In addition, PBH binaries model alone are strongly disfavoured by currently available GW detections, while the best fit was obtained from the template combining ABH and PBH merging models~\citep{Luca2021,Gert2021}. For the sake of a comprehensive comparison, we take the $\beta_{\rm PBH}=0,0.5,1$ in the following analysis. For the mass distribution of BBHs, we take $(m_{\rm c},\sigma)=(30~M_{\odot},0.3)$ as~\citet{Ng2022} work in the PBH binaries, which encompasses the most sensitive mass range of the next-generation GW detectors and can account for the current GW data as presented in~\citet{Chen2019,Wu2020,Luca2020,Luca2021,Gert2021,Kaze2021}. For the mass distribution of ABH binaries, we assume that the mass of black hole is in the range of $5~M_{\odot}\leq m_2<m_1\leq100~M_{\odot}$. The heavier black hole $m_1$ satisfies the mass distribution as equation~(\ref{eq17}), and the mass of $m_2$ uniformly distributes in the interval $[5~M_{\odot},m_1]$. 

\subsection{Bayesian analysis}\label{sec32}
\begin{figure}
    \centering
     \includegraphics[width=0.45\textwidth, height=0.4\textwidth]{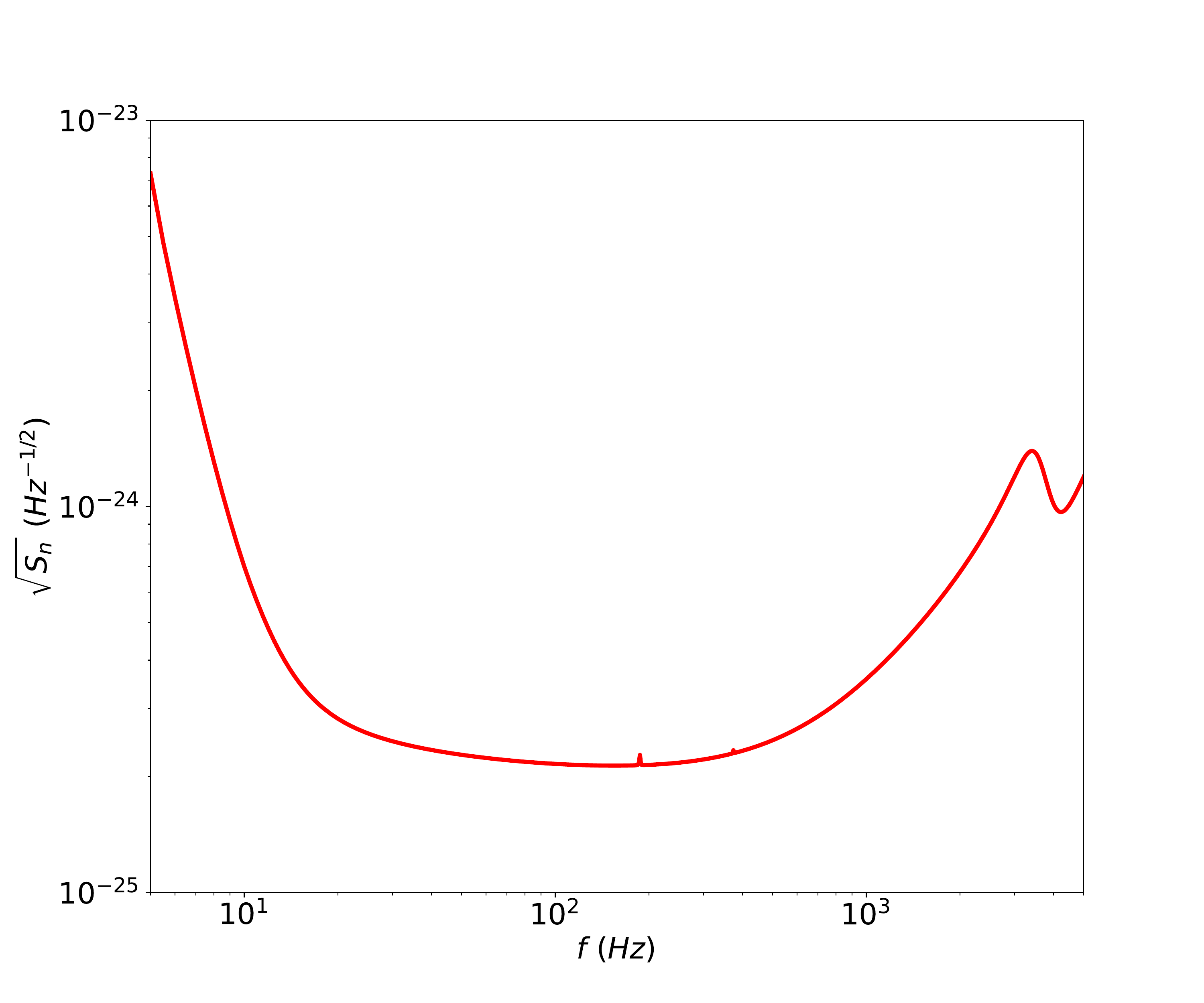}
     \caption{The sensitivity curve of the third-generation GW detector, i.e. the Cosmic Explorer. }\label{fig2}
\end{figure}

\begin{figure*}
    \centering
     \includegraphics[width=0.3\textwidth, height=0.3\textwidth]{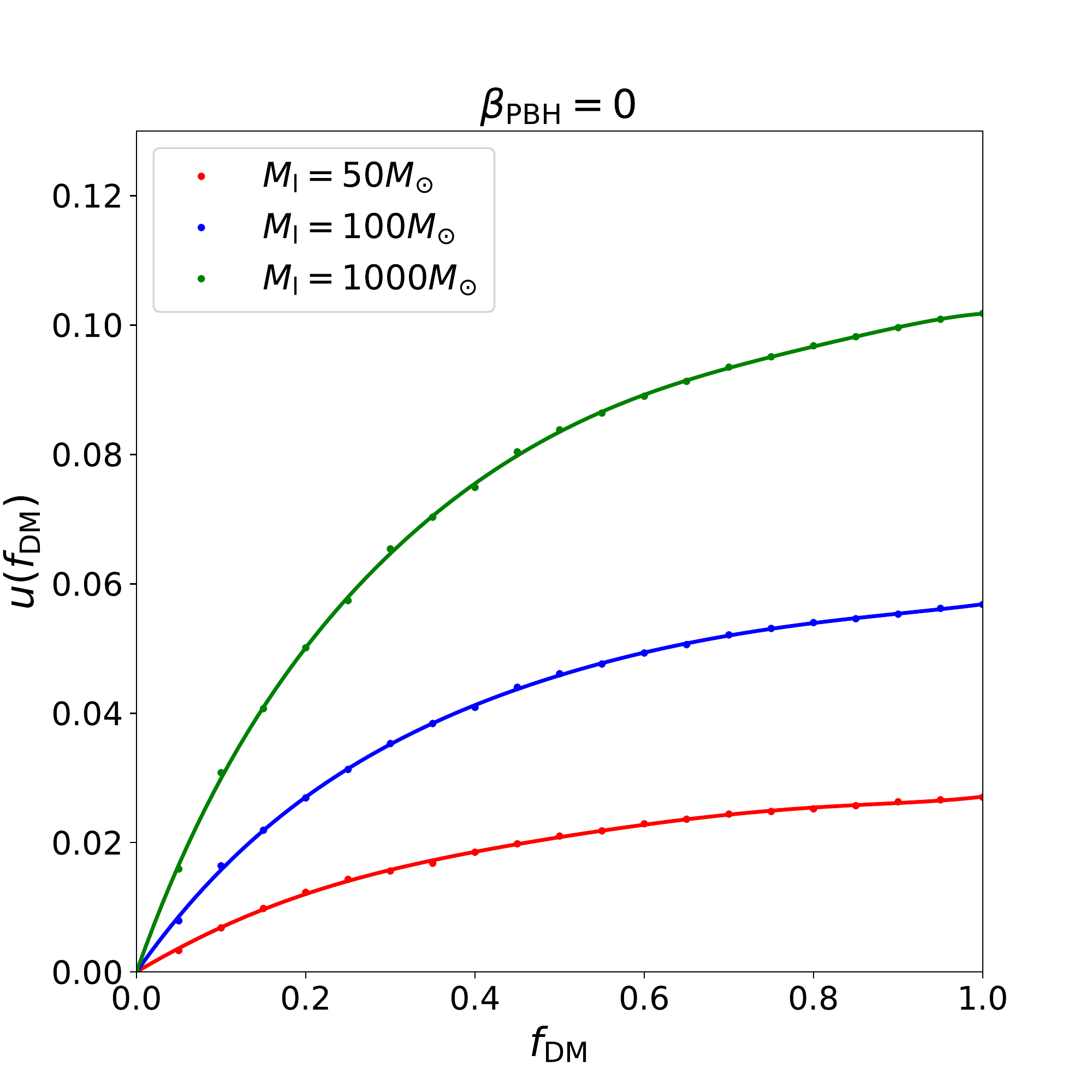}
     \includegraphics[width=0.3\textwidth, height=0.3\textwidth]{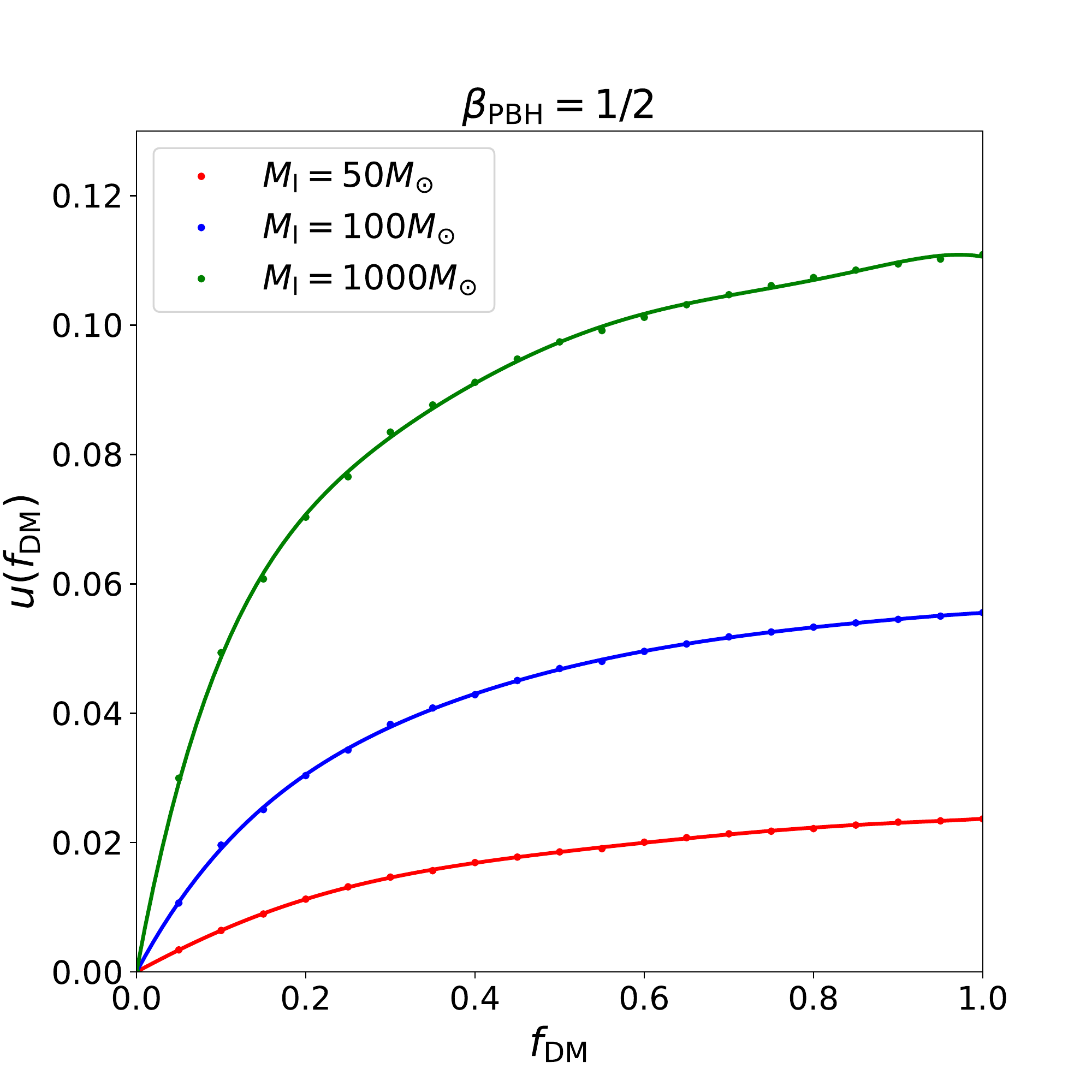}
     \includegraphics[width=0.3\textwidth, height=0.3\textwidth]{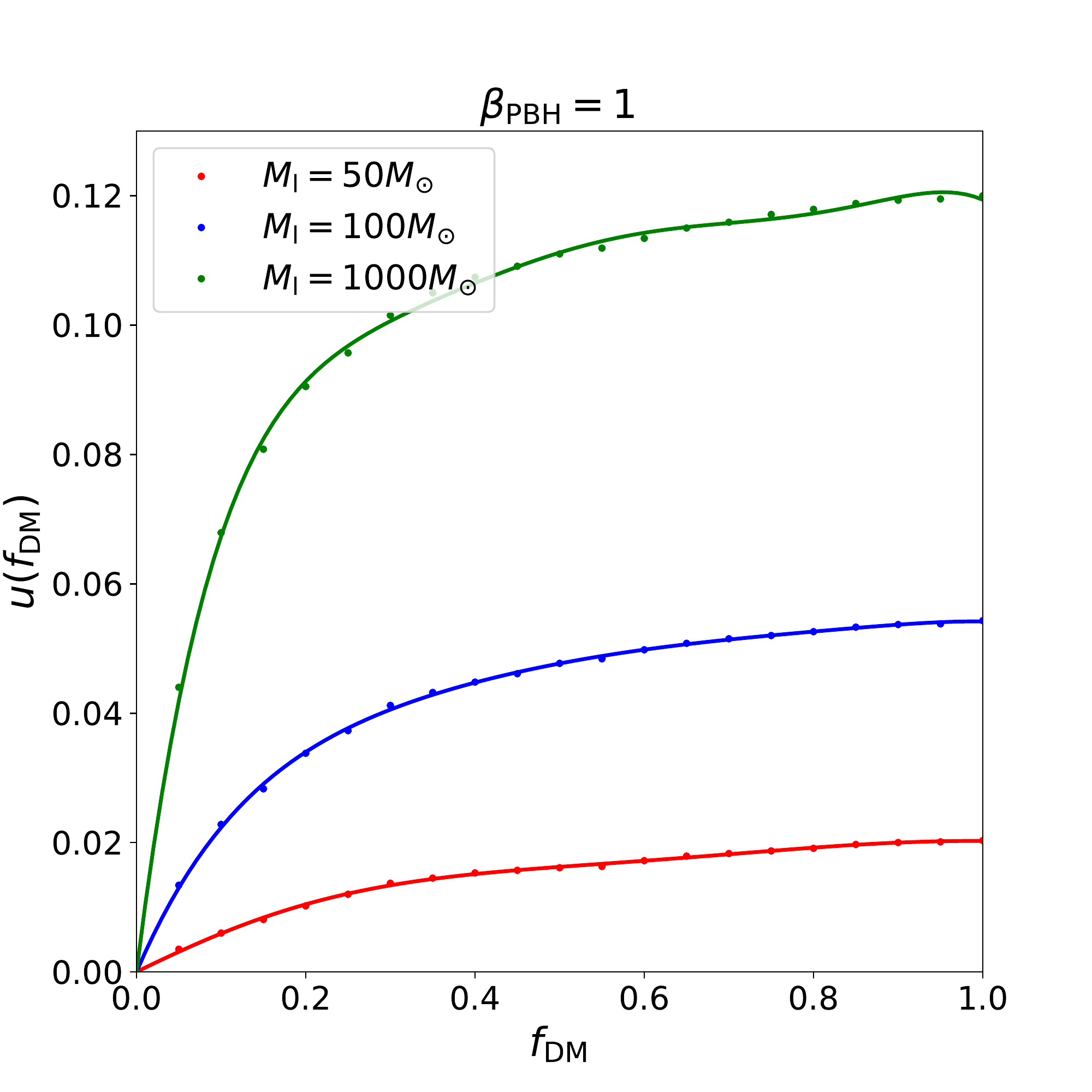}
     \caption{The fraction of simulated events as the process in Section~\ref{sec32}, shown as a function of $f_{\rm DM}$. The left, middle and right subfigures correspond to different $\beta_{\rm PBH}$. In each subfigure, different colors correspond to different lens masses. The curves are power series polynomial fits. }\label{fig3}
\end{figure*}

\begin{figure*}
    \centering
     \includegraphics[width=0.3\textwidth, height=0.3\textwidth]{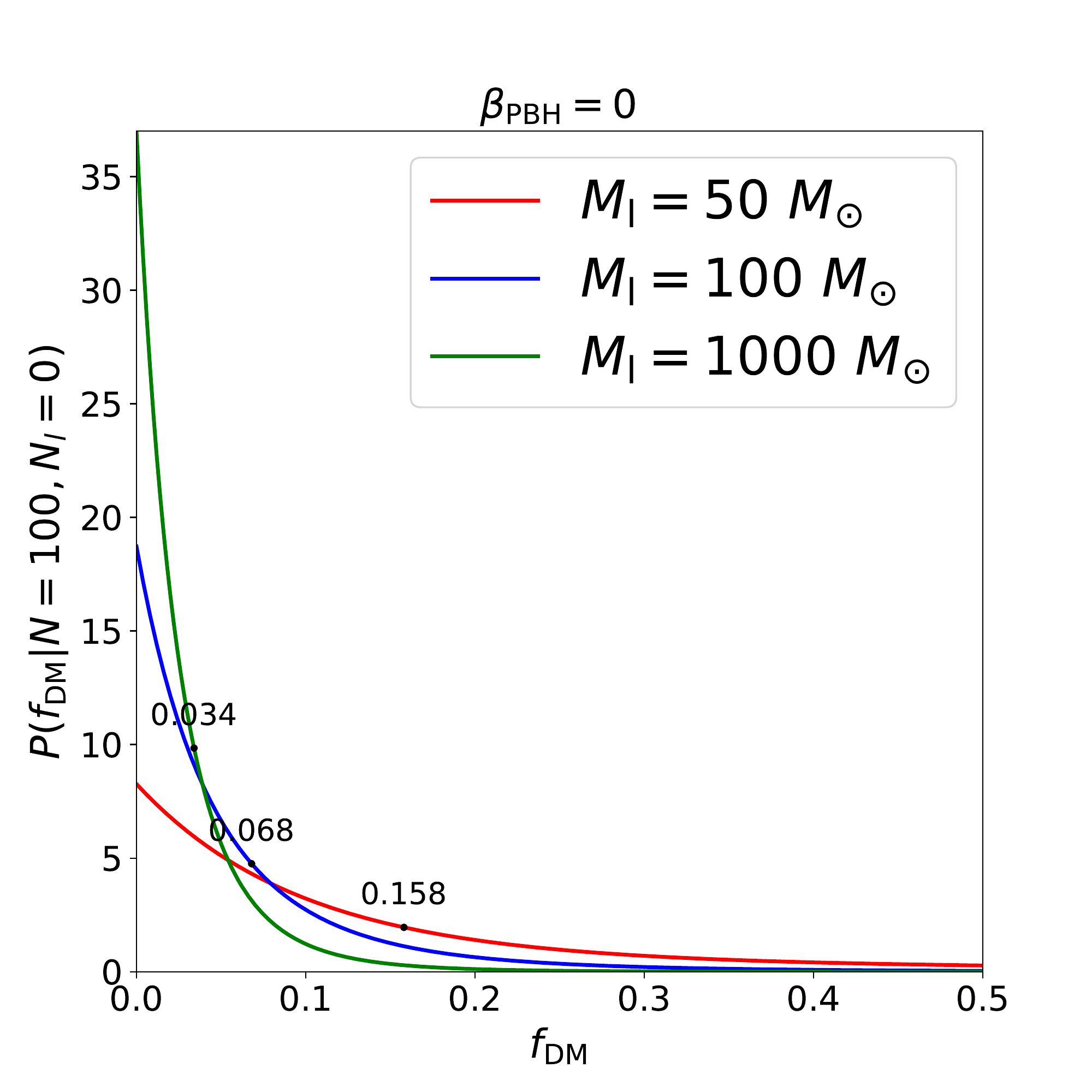}
     \includegraphics[width=0.3\textwidth, height=0.3\textwidth]{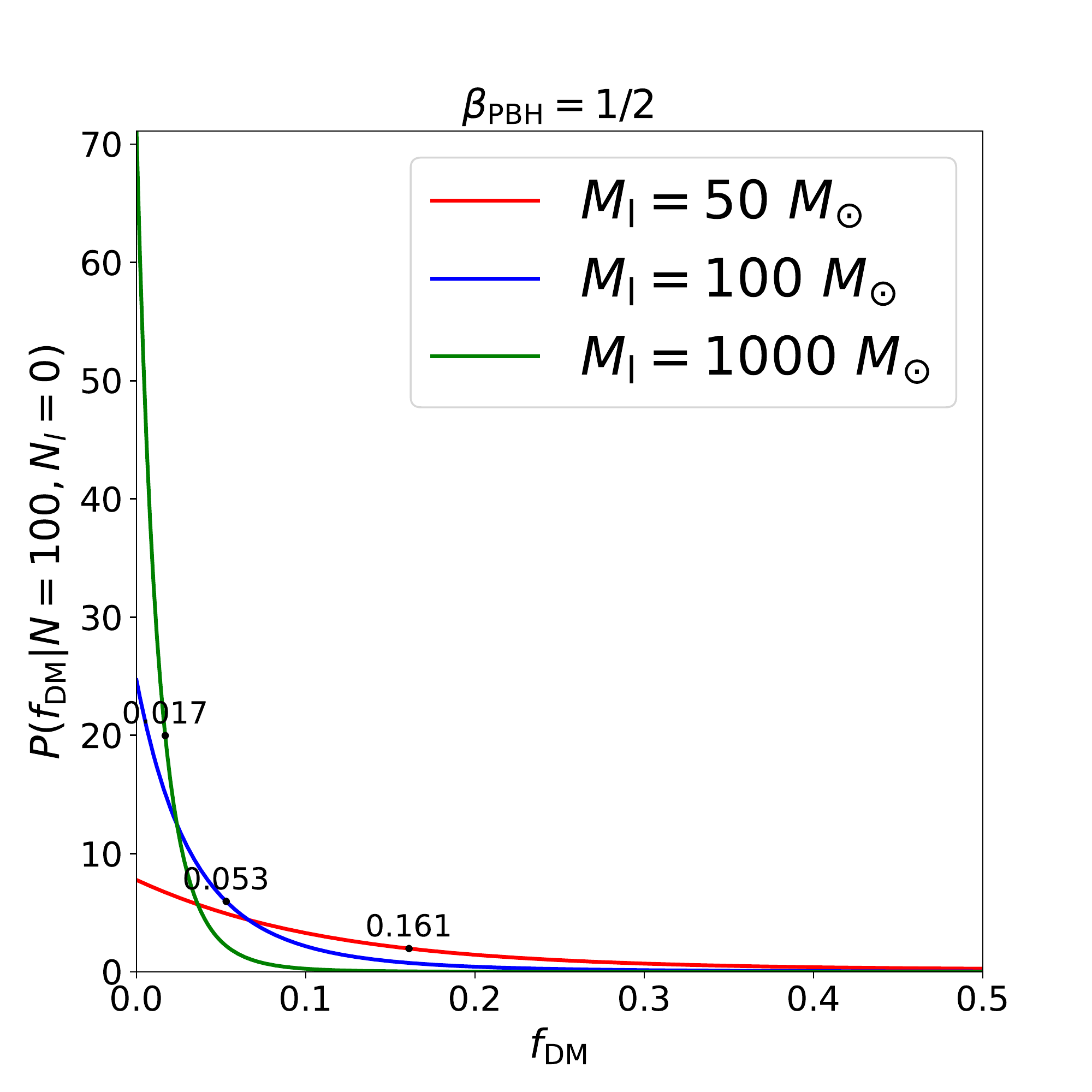}
     \includegraphics[width=0.3\textwidth, height=0.3\textwidth]{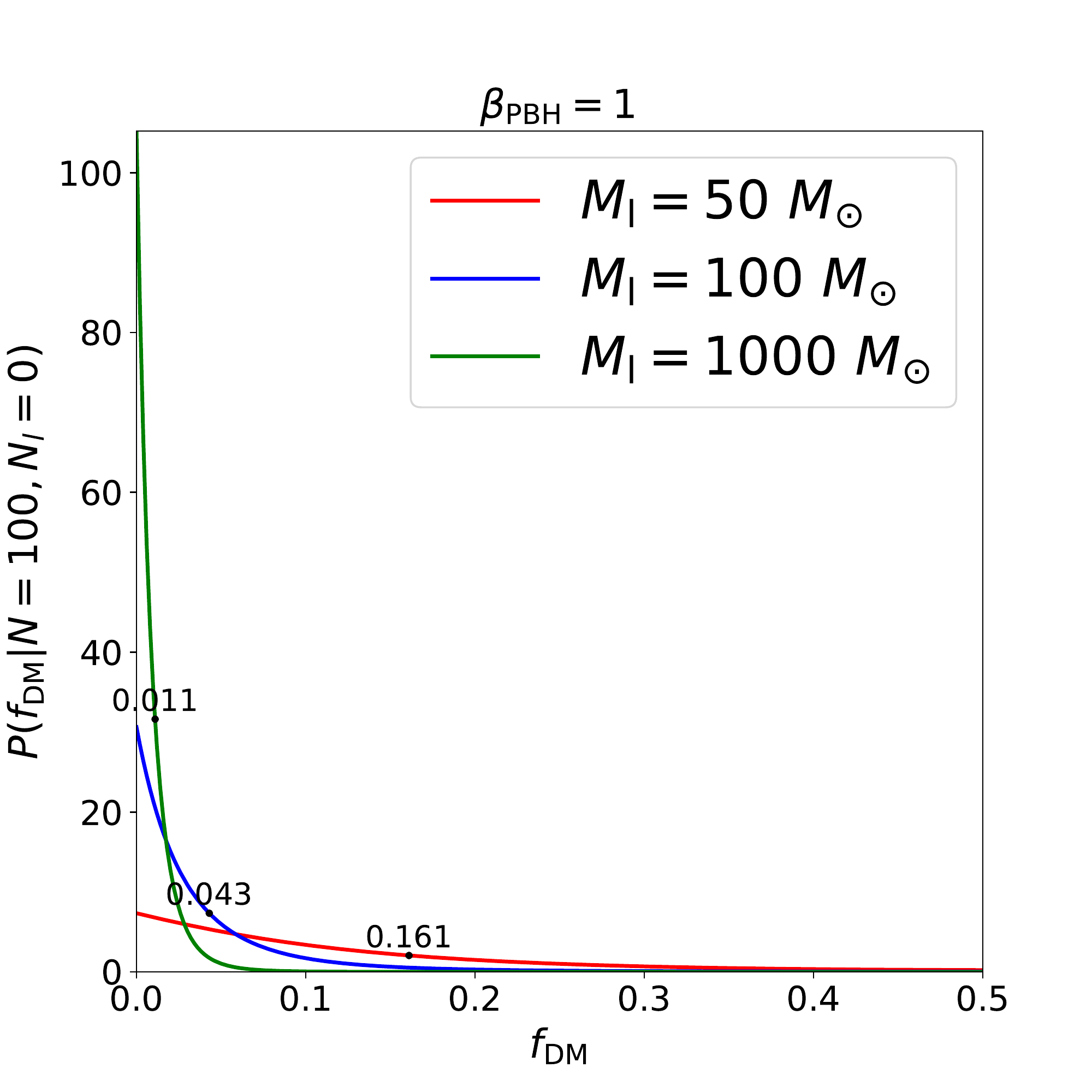}
     \caption{Posteriors on $f_{\rm DM}$ obtained from the non-observation of microlensing signals $N_{\rm l}=0$ in the 100 BBHs events detected by CE. The same as Figure~\ref{fig3}, the left, middle and right correspond to different $\beta_{\rm PBH}$, and curves with different color correspond to different lens masses. The $68\%$ credible upper limits are shown by dots.}\label{fig4}
\end{figure*}

Following the method presented in~\citet{Basak2022}, we introduce the posterior distribution that the mass of lens is lager than $50~M_{\odot}$ to constrain on the $f_{\rm DM}$. We assume that the number of detected GW events of BBH ($N$) follow a Poisson distribution with mean $\Lambda$, whose posterior distribution can be estimated as 
\begin{equation}\label{eq19}
p(\Lambda|N)=\frac{1}{Z}p(\Lambda)p(N|\Lambda),
\end{equation}
where $Z$ is the normalization constant. $p(\Lambda)$ is the flat prior distribution for $\Lambda$
\begin{equation}\label{eq20}
p(\Lambda)=\frac{1}{\Lambda^{\rm max}}\mathcal{H}(\Lambda)\mathcal{H}(\Lambda^{\rm max}-\Lambda),
\end{equation}
where $\Lambda^{\rm max}$ is maximum possible values of $\Lambda$. The likelihood $p(N|\Lambda)$ is approximated by a Poisson distribution 
\begin{equation}\label{eq21}
p(N|\Lambda)=\frac{\Lambda^N\exp(-\Lambda)}{N!}.
\end{equation}
Similarly, from the lensed observations ($N_{\rm l}$), the posterior on the Poisson mean $\Lambda_{\rm l}$ of the number for lensed events can be calculated as
\begin{equation}\label{eq22}
p(\Lambda_{\rm l}|N_{\rm l})=\frac{1}{Z_{\rm l}}p_{\rm l}(\Lambda_{\rm l})p_{\rm l}(N_{\rm l}|\Lambda_{\rm l}),
\end{equation}
where $Z_{\rm l}$ is the normalization factor. we also take $p_{\rm l}(\Lambda_{\rm l})$ as the flat prior distribution for $\Lambda_{\rm l}$
\begin{equation}\label{eq23}
p_{\rm l}(\Lambda_{\rm l})=\frac{1}{\Lambda_{\rm l}^{\rm max}}\mathcal{H}(\Lambda_{\rm l})\mathcal{H}(\Lambda_{\rm l}^{\rm max}-\Lambda_{\rm l}),
\end{equation}
where $\Lambda_{\rm l}^{\rm max}$ is the largest value that corresponds to the situation of $f_{\rm DM}=1$ following the equation (A1) in~\citet{Basak2022}. The likelihood $p(N_{\rm l}|\Lambda_{\rm l})$ is 
\begin{equation}\label{eq24}
p(N_{\rm l}|\Lambda_{\rm l})=\frac{\Lambda_{\rm l}^{N_{\rm l}}\exp(-\Lambda_{\rm l})}{N_{\rm l}!}.
\end{equation}
To calculate the posterior on the fraction of lensed events $u\equiv\Lambda_{\rm l}/\Lambda$, we can use the ratio distribution assuming the $\Lambda$ and $\Lambda_{\rm l}$ are independent to get
\begin{equation}\label{eq25}
p(u|\{N,N_{\rm l}\})=\frac{1}{Z_{u}}\int_0^{+\infty}\Lambda p(\Lambda|N)p(u\Lambda|N_{\rm l})d\Lambda,
\end{equation}
where $Z_{u}$ is normalization constant that determined by $\int_0^{u^{\rm max}}p(u|\{N,N_{\rm l}\})du=1$, where $u^{\rm max}$ is maximum value of $u$ (corresponding to $f_{\rm DM}=1$). Finally, the posterior of $f_{\rm DM}$ can be writen as
\begin{equation}\label{eq26}
p(f_{\rm DM}|\{N,N_{\rm l}\})=\frac{1}{Z_{f_{\rm DM}}}p(u|\{N,N_{\rm l}\})\bigg|\frac{du}{df_{\rm DM}}\bigg|,
\end{equation}
where $Z_{f_{\rm DM}}$ is normalization constant and $\bigg|\frac{du}{df_{\rm DM}}\bigg|$ is the Jacobian determinant. We determine this Jacobian determinant as these steps:

\begin{itemize}
\item\textbf{Generating the detectable BBH events:} Based on the redshift distribution and mass distribution in Simulation subsection, we can simulate $\mathcal{O}(10^4)$ detectable unlensed BBH events with signal-to-nosie ratio ($\rm{SNR}\geq8$) for CE. SNR of unlensed GW can be computed as
\begin{equation}\label{eq27}
{\rm{SNR}}=\sqrt{4\int^{f_1}_{f_0}\frac{|\bar{h}_0(f)|^2}{S_{\rm n}(f)}df},
\end{equation}
where the lower cutoff $f_0$ depends on the sensitivity curve  $\sqrt{S_{\rm n}(f)}$ of CE~\footnote{Sensitivity curve of CE is available via https://cosmicexplorer.org/sensitivity.html} as shown in Figure~\ref{fig2}, and we adopt $f_0=5~\rm{Hz}$. The cutoff frequency $f_1$ is adopted by the value ${\rm{min}}\{[3\sqrt{3}\pi(1+z_{\rm s})(m_1+m_2)]^{-1},5~\rm{kHz}\}$.

\item\textbf{Generating the lensed GW events:} By assuming that the compact DM is distributed uniformly in comving volume, the probability that a GW source at $z_{\rm s}$ is lensed by an intervening compact DM object can be given by $P_{\rm l}(z_{\rm s})=1-\exp(-\tau(z_{\rm s}))$. We identify a BBH GW signal as a lensed event when $P_{\rm l}(z_{\rm s})$ is larger than a random number uniformly distribution between 0 and 1. Therein the optical depth $\tau(z_{\rm s})$ can be written as
\begin{equation}\label{eq28}
\tau(z_{\rm s})=\int_0^{z_{\rm s}}dz_{\rm l}\int_0^{y_0}dy~\tau(z_{\rm s},z_{\rm l},y),
\end{equation}
where $y_0$ is the maximum impact parameter. $\tau(z_{\rm s},z_{\rm l},y)$ is the differential optical depth for the redshift of lens $z_{\rm l}$ and impact parameter $y$~\citep{Liao2020a,Urrutia2021,Wang2021,Basak2022}
\begin{equation}\label{eq29}
\tau(z_{\rm s},z_{\rm l}, y)=3f_{\rm DM}\Omega_{\rm DM}y\frac{H_0^2(1+z_{\rm l})^2}{H(z_{\rm l})}\frac{D_{\rm l}D_{\rm ls}}{D_{\rm s}}.
\end{equation}
When a GW event is identified as a lensed one, the $z_{\rm l}$ is adopted from a probability distribution $P(z_l)$ given by equation~(\ref{eq29})
\begin{equation}\label{eq30}
P(z_l)=\frac{\tau(z_{\rm s},z_{\rm l},y)}{\int_0^{z_{\rm s}}dz_{\rm l}~\tau(z_{\rm s},z_{\rm l},y)}.
\end{equation}
In addition, the distribution $P(y)$ also can be obtained from equation~(\ref{eq29}) in the range of $y\in[0,y_0]$
\begin{equation}\label{eq31}
P(y)=\frac{\tau(z_{\rm s},z_{\rm l},y)}{\int_0^{y_0}dy~\tau(z_{\rm s},z_{\rm l},y)}=\frac{2}{y_0^2}y.
\end{equation}
We set $y_0=5$ since GW events with $y>5$ are difficult to be identified as lensing signals~\footnote{Larger values of $y_0$ would lead to weaker constraints on $f_{\rm DM}$.}.

\item\textbf{Identifying BBH events with wave optics effect:} The time delay caused by a point mass is given by
\begin{equation}\label{eq32}
\Delta t_{\rm lening}=4M_{\rm l}^z\bigg[\frac{y\sqrt{y^2+4}}{2}+\ln\bigg(\frac{\sqrt{y^2+4}+y}{\sqrt{y^2+4}-y}\bigg)\bigg],
\end{equation}
The duration of GW signal can be approximated as 
\begin{equation}\label{eq33}
t_{\rm signal}=\frac{5}{256}\mathcal{M}_{\rm s}^{z-5/3}(\pi f_0)^{-8/3}+10^4M_{\rm s}^z,
\end{equation}
where $\mathcal{M}_{\rm s}^z$ and $M_{\rm s}^z$ are redshifted chirp mass and total mass of BBH, respectively. We consider those lensed signals with $\Delta t_{\rm lening}\leq t_{\rm signal}$ satisfy wave optics effects. 
\item\textbf{Computing lensed criterion:} After generating unlensed and lensed gravitational waveform with simulated source and lens parameters, we use the test SNR as criterion following~\citet{Jung2019,Liao2020a}
\begin{equation}\label{eq34}
{\rm{SNR_{test}}}=\sqrt{4\int^{f_1}_{f_0}\frac{|\bar{h}^{\rm L}(f)-\bar{h}_{\rm best-fit}(f)|^2}{S_{\rm n}(f)}df}.
\end{equation}
To ensure that the difference between the lensed GW signal and the unlensed signal is significant enough, which leads to figure out the lensing signal, we adopt $\rm{SNR_{test}}\geq8$ as the lensed criterion. In addition, we use the standard template to fit the lensed signal by varying the amplitude and phase.

\item\textbf{Calculating $u(f_{\rm DM})$:} We can obtain $u(f_{\rm DM})$ 
\begin{equation}\label{eq35}
u(f_{\rm DM})=\frac{N_{\rm lensed}(f_{\rm DM})}{N_{\rm t}},
\end{equation}
where $N_{\rm lensed}$ and $N_{\rm t}$ represent the number of lensed events expected to be detected and total number of detectable events, respectively. If we take the mass of lens $M_{\rm l}$ as a constant, $u(f_{\rm DM})$ corresponds to the fraction of simulated events  for the monochromatic mass distribution of lens. Here we also take a extend mass function for the lens with the power-law function as~\cite{Laha2020} work
\begin{equation}\label{eq36}
P_{\rm l}(M_{\rm l})=\mathcal{N}_{\rm pl}M_{\rm l}^{\gamma-1}\mathcal{H}(M_{\rm l}-M_{\min})\mathcal{H}(M_{\max}-M_{\rm l}),
\end{equation}
where the mass range of the distribution is bordered by the minimum mass, $M_{\min}$, and the maximum mass, $M_{\max}$. The exponent of the power law is denoted by $\gamma$. $\mathcal{N}_{\rm pl}$ is the normalization constant as 
\begin{equation}\label{eq37}
\mathcal{N}_{\rm pl}=\left\{
\begin{aligned}
\frac{\gamma}{M_{\max}^{\gamma}-M_{\min}^{\gamma}},\quad\gamma\neq0,\\
\frac{1}{\ln(M_{\max}/M_{\min})},\quad\gamma=0,
\end{aligned}
\right.
\end{equation}

\end{itemize}
Then we can obtain the limit of $f_{\rm DM}$ from the posterior distribution following Equation~(\ref{eq26}).

In addition to the Bayesian analysis method, we also use the optical depth method in our results for comparison. For this method, we can use equation~(\ref{eq34}) to obtain the maximum impact parameter $y_{\rm max}$, and then calculate the cross section of the lens
\begin{equation}\label{eq38}
\sigma(M_{\rm l},z_{\rm s},z_{\rm l})=4\pi M_{\rm l}\frac{D_{\rm l}D_{\rm ls}}{D_{\rm s}}y_{\rm max}^2.
\end{equation}
With the above cross section, we can further get the optical depth $\tau_i$ for each GW event
\begin{equation}\label{eq39}
\begin{split}
\tau_i(M_{\rm l},f_{\rm DM},z_{\rm s})=\int_0^{z_{\rm s}}d\chi(z_{\rm l})(1+z_{\rm l})^2n_{\rm l}(f_{\rm DM})\sigma(M_{\rm l},z_{\rm s},z_{\rm l})\\
=\int_0^{z_{\rm s}}dz_{\rm l}\frac{3}{2}f_{\rm DM}\Omega_{\rm DM}y^2_{\max}\frac{H_0^2(1+z_{\rm l})^2}{H(z_{\rm l})}
\frac{D_{\rm l}D_{\rm ls}}{D_{\rm s}}.
\end{split}
\end{equation}
According to the Poisson law, the probability of failing to detect lensed event is  
\begin{equation}\label{eq40}
P_i=\exp(-\tau_i(M_{\rm l},f_{\rm DM},z_{\rm s})).
\end{equation}
If we have detected a large number of GW evnets $N_{\rm total}$, and none of them has been lensed, the total probability of unlensed event would be given by
\begin{equation}\label{eq41}
P_{\rm tot}=\exp(-\sum^{N_{\rm total}}_{i=1}\tau_i).
\end{equation}
If none lensed detection is consistent with the hypothesis that the universe is filled with the compact DM to a fraction $f_{\rm DM}$ at $100\Pi\%$ confidence level, the following condition must be valid
\begin{equation}\label{eq42}
P_{\rm tot}(f_{\rm DM})\geq1-\Pi.
\end{equation}
For a null search of lensed GW signals, then the constraint on the upper limit of $f_{\rm DM}$ can be estimated from equation~(\ref{eq42}). For the optical depth $\tau_i\ll1$ , we can obtain the expected number of lensed event from a large number of GW evnets
\begin{equation}\label{eq43}
N_{\rm lensed}(M_{\rm l}, f_{\rm DM})=\sum^{N_{\rm total}}_{i=1}(1-\exp(-\tau_i))\approx\sum^{N_{\rm total}}_{i=1}\tau_i.
\end{equation}

\begin{figure}
    \centering
     \includegraphics[width=0.45\textwidth, height=0.45\textwidth]{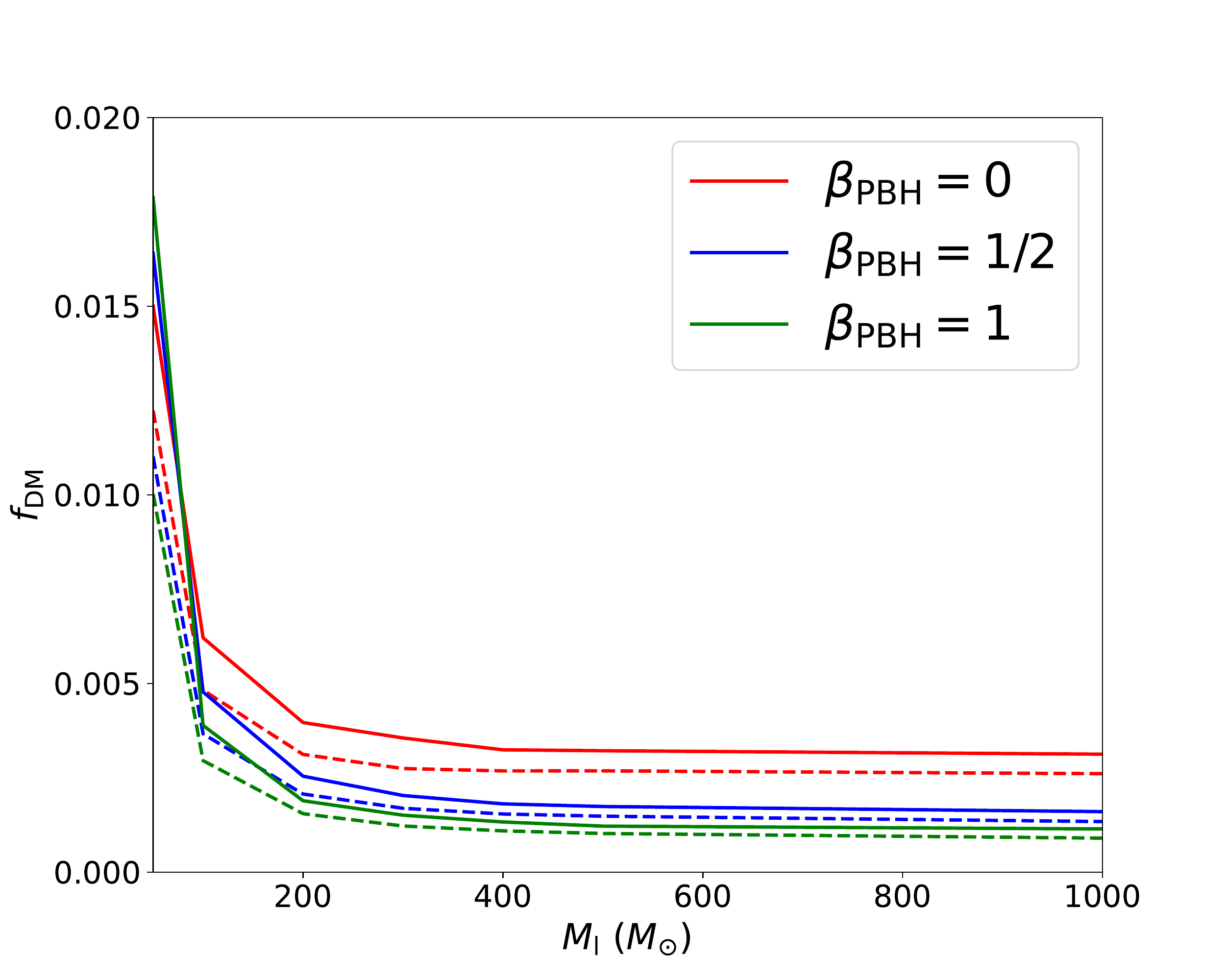}
     \caption{Constraints on the upper limits of fraction of dark matter in the form of compact DM at $68\%$ confidence level, shown as a function of lens mass with 1000 observed BBH events. Lines with different color correspond to results with different $\beta_{\rm PBH}$. Solid and dashed lines represent the upper limits using the Bayesian analysis and the optical depth, respectively.}\label{fig5}
\end{figure}

\begin{figure}
    \centering
     \includegraphics[width=0.45\textwidth, height=0.45\textwidth]{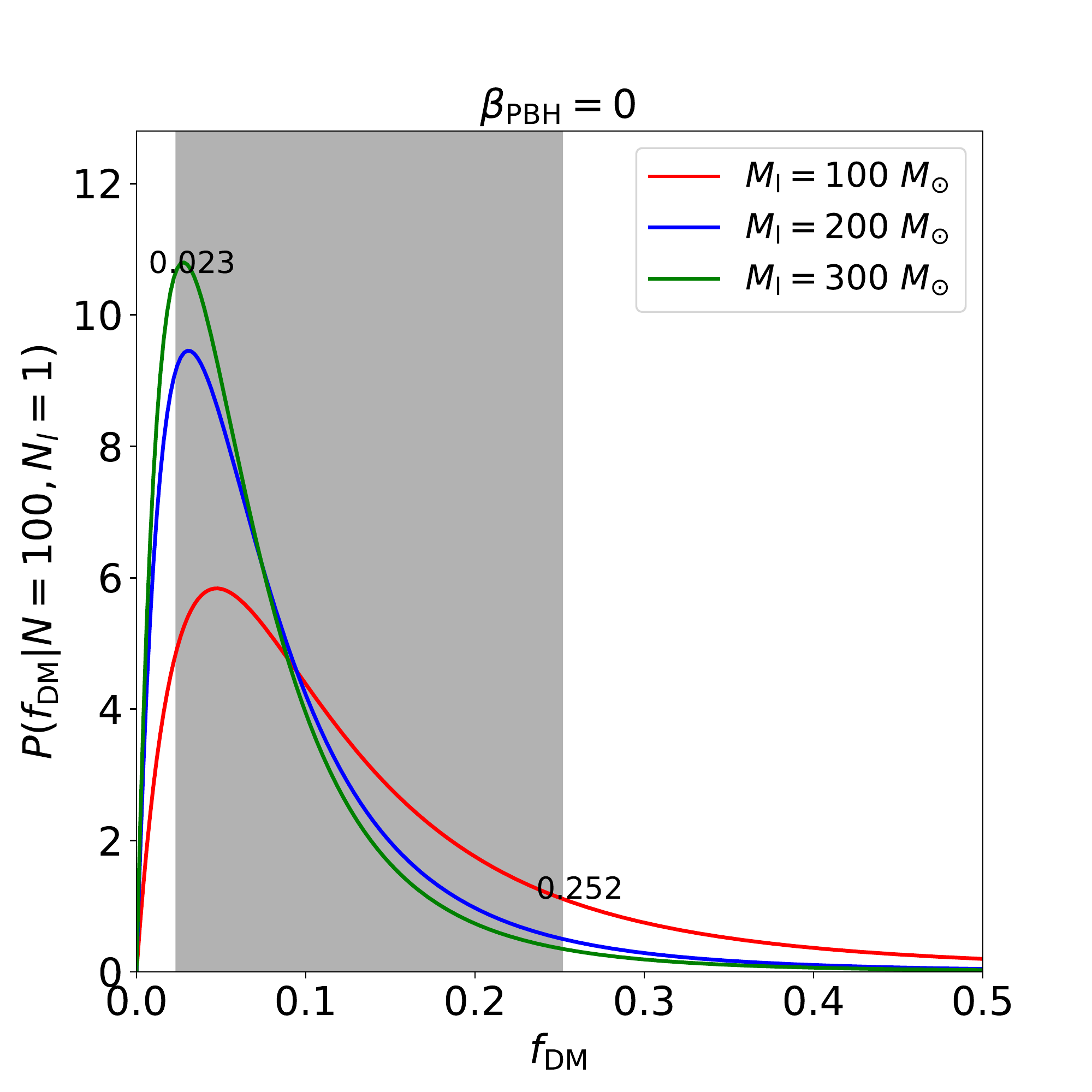}
     \caption{Similar as Figure~\ref{fig4}, posteriors on $f_{\rm DM}$ obtained from an identification of microlensing signals $N_{\rm l}=1$ in the 100 BBHs events detected by CE. Lines with different color correspond to results with different different lens masses. The shadow represents the range of $f_{\rm DM}$ with the redshift of source being 2 and the redshifted lens mass being $300~M_{\odot}$.}\label{fig6}
\end{figure}

\begin{figure*}
    \centering
     \includegraphics[width=0.45\textwidth, height=0.45\textwidth]{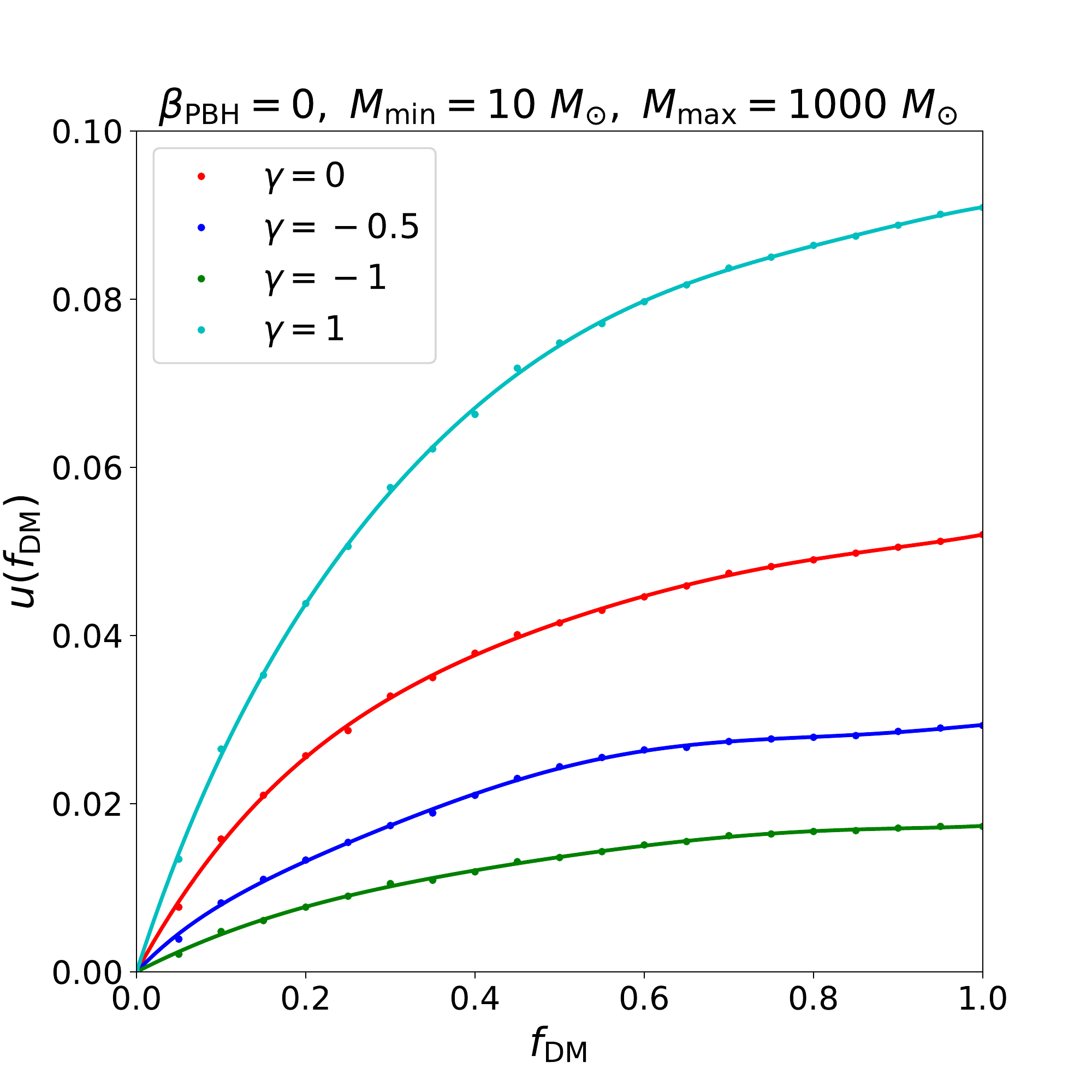}
     \includegraphics[width=0.45\textwidth, height=0.45\textwidth]{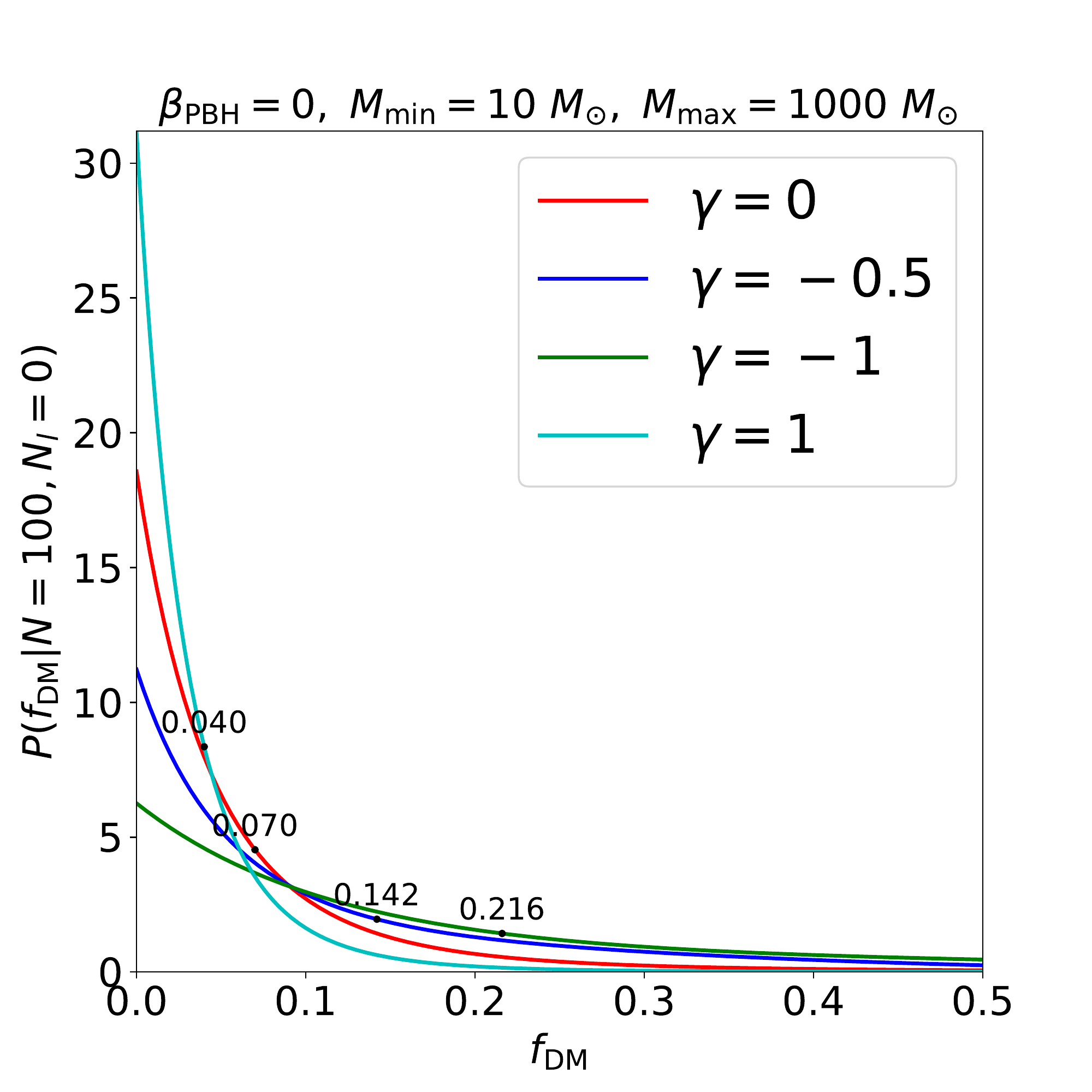}
     \caption{The same as Figure~\ref{fig3}, left panel shows the fraction of simulated events with a power law mass function for lens. Right panel shows posteriors with $68\%$ confidence level. Curves with different color correspond to results with different power exponent $\gamma=-1, -0.5, 0, 1$. }\label{fig7}
\end{figure*}

\section{Results}\label{sec4}
We consider two formation channels of BBHs, i.e. the ABH binary scenario and the PBH binary scenario, to contribute the low and high redshift GW events with a parameterized fraction $\beta_{\rm PBH}$. For the monochromatic mass distribution of lens mass, the fraction of simulated events with $\rm{SNR_{test}}\geq8$ is shown as a function of $f_{\rm DM}$ in Figure~\ref{fig3}. This enables us to calculate the posterior of  $f_{\rm DM}$. Firstly, Figure~\ref{fig4} shows the posterior of $f_{\rm DM}$ assuming 100 observed BBH events without lensed signal identification. The $68\%$ upper limits are shown as black dots in each subfigure of Figure~\ref{fig4}. We find that $\beta_{\rm PBH}$ is larger (more high redshift GW events of BBH), the constraints on $f_{\rm DM}$ are stronger. However, there is no significant improvement of these constraints. In Figure~\ref{fig5}, we show the constraints on $f_{\rm DM}$ from the detection of 1000 BBH events without lensed signal identification by using the Bayesian analysis and the optical depth, respectively. We find that the constraints on $f_{\rm DM}$ of Bayesian analysis can be $0.32\%$ ($\beta_{\rm PBH}=0$), $0.16\%$ ($\beta_{\rm PBH}=1/2$), $0.12\%$ ($\beta_{\rm PBH}=1$) for $\geq500~M_{\odot}$ at $68\%$ confidence level, respectively. In addition, corresponding constraints using the optical depth method are $0.26\%$ ($\beta_{\rm PBH}=0$), $0.14\%$ ($\beta_{\rm PBH}=1/2$), $0.10\%$ ($\beta_{\rm PBH}=1$), respectively. Although there is tiny difference between results obtained with these two methods, the magnitudes of them are well consistent. This tiny difference may be due to a few factors, e.g. amount of simulated data, prior distributions on the $\Lambda$ and $\Lambda_{\rm l}$, the independence of $\Lambda$ and $\Lambda_{\rm l}$. The fraction of DM made up of compact DM could be constrained to $\sim0.1\%-0.3\%$ for $\geq500~M_{\odot}$ at $68\%$ confidence level, which is improved by an order of magnitude compared with the results of 100 observed BBH events. However, our constraints of $f_{\rm DM}$ based on the assumption of observed 1000 GW events are slightly conservative. \citet{Chen2020} predicted that the third-generation GW detectors like CE are expected to detect $\mathcal{O}(10^5)$ and $\mathcal{O}(10^4)$ BBH mergers each year for PBH binaries model and ABH binaries model, respectively. Therefore, future GW observations would be able to yield stronger constraints on $f_{\rm DM}$. In addition to the null search of lensed GWs, we also assume the situation ($\beta_{\rm PBH}=0$) that a BBH source with redshift $z_{\rm s}=2$ is lensed by a compact DM object with the redshifted lens mass of $300~M_{\odot}$. In this case, the lens mass would be in the range of $[100~M_{\odot},300~M_{\odot}]$. As shown in the Figure~\ref{fig6}, we can obtain $68\%$ confidence level of the $f_{\rm DM}$ from the posterior distribution at this mass range. The constraints on $f_{\rm DM}$ are $0.105_{-0.064}^{+0.147}$ with $100~M_{\odot}$, $0.064_{-0.039}^{+0.083}$ with $200~M_{\odot}$ and $0.057_{-0.034}^{+0.071}$ with $300~M_{\odot}$ at $68\%$ confidence level. Therefore if the lens mass is uniformly distributes in $[100~M_{\odot},300~M_{\odot}]$, the constraints on the $f_{\rm DM}$ can be between $2.3\%$ and $25.2\%$. 

For extended mass distribution of lens mass, we assume that the maximum value of the extended power-law mass function $M_{\max}=10^3~M_{\odot}$ and minimum value $M_{\min}=10~M_{\odot}$. In addition, we assume a power-law mass function with different power exponent $\gamma=-1, -0.5, 0, 1$. The power exponent $\gamma=1$ corresponds to the uniform distribution between the the minimum mass $M_{\min}$ and  maximum mass $M_{\max}$. Here, we just analyze the results for $\beta_{\rm PBH}=0$. Then the fraction of simulated events and the posterior of $f_{\rm DM}$ assuming 100 detected BBH events with null lensed search are shown in Figure~\ref{fig7}. In this model, the $\gamma=1$ case corresponds to more flat mass distribution, which leads more lens of larger mass to contribute lensing probability and stronger limits. Intuitively,~\citet{Carr2017} proposed a simple formula for applying constraints with the monochromatic mass distribution to specific extended mass function as 
\begin{equation}\label{eq45}
\int dM_{\rm l} \frac{f_{\rm DM}P_{\rm l}(M_{\rm l})}{f_{\max}(M_{\rm l})}\leq1.
\end{equation}
where $f_{\max}(M_{\rm l})$ is the maximally allowed value of $f_{\rm DM}$ from the monochromatic mass distribution. Here, for the~\citet{Carr2017} method with $\gamma=1$ and $N=100$, we obtain constraints on the $f_{\rm DM}$ can be derived at $0.039$, which is consistent with the result presented in Figure~\ref{fig7}. Later, this method has also been developed by~\citet{Bellomo2018}. They proposed that constraints with any specific mass function can be related to results from monochromatic mass distribution by the equivalent mass. For instance, comparing with the results shown in the left plot of Figure~\ref{fig4}, we find that the equivalent mass for the power-law mass function with $\gamma=0$ is approximately equal to $100~M_{\odot}$.

\section{Conclusion}\label{sec5}
In this paper, we have derived constraints on the $f_{\rm DM}$ at mass range $\geq50~M_{\odot}$ by assuming two representative scenarios of GW sources. First, we apply the method of Bayesian analysis for CE, a future third-generation ground-based GW detector, to derive constraints on the abundance of compact DM, $f_{\rm DM}$. We obtain that it can be constrained to $\leq1\%$ for $\geq500~M_{\odot}$ at $68\%$ confidence level from 100 unlensed GW events of BBH. If one lensed signal is identified from 1000 GW events, the constraint on $f_{\rm DM}$ will be enhanced by an order of magnitude. This result can be compared with other observational limitations in the future, such as lensing effect of FRB~\citep{Zhou2022a}. In addition, if there are 100 observed BBH events from CE in the future, we find that a BBH GW events with redshift $z_{\rm s}=2$ is lensed by a compact DM object with the redshifted lens mass $M_{\rm l}^{z}=300~M_{\odot}$. Then the $f_{\rm DM}$ sholud be constrained to between $2.3\%$ and $25.2\%$. In addition to the above constraints on the abundance of compact DM in the framework of monochromatic mass distribution of lens, we also consider a test extended mass distribution of lens to get more general constraints of $f_{\rm DM}$. After deducing the equivalent mass for a given input of extended mass distribution, we can then read off the $f_{\rm DM}$ for that distribution by using Figure~\ref{fig4} and Figure~\ref{fig7}. It should be noted that our analysis neglects the macrolensing effect from the structures in hosting galaxy~\citep{Diego2020,Cheung2021}. Besides, possible astrophysical microlens e.g. stellar, in hosting galaxies may contaminate the constraints on the abundance of compact DM~\citep{Christian2018}.

Constraints on the $f_{\rm DM}$ will be significantly improved because of the rapid increase of the number of GWs detected by various surveys and increased horizon distance of the next generation GW detectors in the near future. As a result, there would be significant overlap between the areas constrained from GW detection and the one from other electromagnetic observations. Then, it will be possible to jointly constrain the abundance and mass distribution of compact DM by combining these two kinds of promising multi-messenger observations. It is foreseen that these joint constraints will be of great importance for exploring the nature of compact DM or even their formation mechanisms relating to the physics of the early universe.

\section*{Acknowledgements}
This work was supported by the National SKA Program of China No. 2020SKA0110402; National key R\&D Program of China (Grant No. 2020YFC2201600); National Key Research and Development Program of China Grant No. 2021YFC2203001; National Natural Science Foundation of China under Grants Nos. 12275021, 11920101003, 11722324, 11603003, 11633001, 12073088, and U1831122; Guangdong Major Project of Basic and Applied Basic Research (Grant No. 2019B030302001), the Strategic Priority Research Program of the Chinese Academy of Sciences, Grant No. XDB23040100, and the Interdiscipline Research Funds of Beijing Normal University. 

\section*{Data Availability}
The data underlying this article will be shared on reasonable request to the corresponding author.

\label{lastpage}
\end{document}